\documentclass[12pt]{revtex4}
\usepackage{epsfig}
\usepackage{amsmath,amsthm,amssymb}
\def\be{\begin{equation}}
\def\ee{\end{equation}}
\def\bea{\begin{eqnarray}}
\def\eea{\end{eqnarray}}
\usepackage{graphicx}

\catcode`\@=11
\def\lsim{\mathrel{\mathpalette\@versim<}}
\def\gsim{\mathrel{\mathpalette\@versim>}}
\def\@versim#1#2{\vcenter{\offinterlineskip
\ialign{$\m@th#1\hfil##\hfil$\crcr#2\crcr\sim\crcr } }}
\catcode`\@=12
\usepackage{axodraw}

\catcode`@=12
\begin{document}
\thispagestyle{empty}
\begin{flushright}
UCRHEP-T468\\
May 2009\
\end{flushright}
\vspace{0.8in}

\begin{center}
{\LARGE \bf Neutrino Mass: Mechanisms and Models\\}
\vspace{1.2in}
{\bf Ernest Ma\\}
\vspace{0.2in}
{\sl Department of Physics and Astronomy, University of California,\\
Riverside, California 92521, USA\\}
\end{center}
\vspace{2.2in}

In these lectures (at the 2007 Summer School in Akyaka, Mugla, Turkey), I 
discuss the various mechanisms for obtaining small Majorana neutrino masses, 
as well as specific models of varying complexity, in the context of the 
standard model and beyond.

\newpage

\section{INTRODUCTION}

In the Standard Model (SM) of particle interactions based on the gauge 
group $SU(3)_C \times SU(2)_L \times U(1)_Y$, all fermions (and gauge bosons) 
owe their masses to its $one$ Higgs scalar doublet $\Phi = (\phi^+,\phi^0)$. 
In particular, the charged-lepton mass comes from
\begin{equation}
f_l (\bar{\nu}_L \phi^+ + \bar{l}_L \phi^0) l_R \Rightarrow m_l = f_l \langle 
\phi^0 \rangle.
\end{equation}
The lone exception is the neutrino because the singlet $\nu_R$ is trivial 
under the SM gauge group, i.e. $\nu_R \sim (1,1,0)$, so it is not required 
to be part of the SM.  Thus the minimal SM has zero neutrino mass, which 
is of course not realistic, in the face of established neutrino-oscillation 
data in the last decade.  In the following lectures, I will discuss the 
generic mechanisms for obtaining small Majorana neutrino masses, and 
specific models which realize them in the context of the SM and beyond.
I will also discuss $A_4$ briefly for understanding tribimaximal neutrino 
mixing.

\subsection{TYPE I SEESAW} 

The most prevalent idea for obtaining a neutrino mass is to add $\nu_R$, then 
\begin{equation}
f_\nu (\bar{\nu}_L \bar{\phi}^0 - \bar{l}_L \phi^-) \nu_R \Rightarrow 
m_D = f_\nu \langle \bar{\phi}^0 \rangle
\end{equation}
is a fermion Dirac mass just like $m_l$.  However, since $\nu_R$ is a gauge 
singlet, it can have a Majorana mass $M$, so that the $2 \times 2$ mass 
matrix linking $\bar{\nu}_L$ to $\nu_R$ is of the form
\begin{equation}
{\cal M}_\nu = \begin{pmatrix}0 & m_D \\ m_D & M \end{pmatrix},
\end{equation}
with eigenvalues $m_{1,2} = M/2 \mp \sqrt{(M/2)^2 + m_D^2}$.  There are two 
interesting limits, as discussed below. 

(a) If $M=0$, then $m_{1,2} = \mp m_D$ and $\nu_L$ pairs with $\nu_R$ to form 
a Dirac fermion with additive lepton number $L=1$, which is exactly conserved. 
This also shows that a neutral Dirac fermion may be regarded as two 
mass-degenerate Majorana fermions of opposite $CP$. It is a perfectly 
acceptable explanation of neutrino mass, but it requires a very tiny 
$f_\nu$ in Eq.~(2), of order $10^{-11}$ or less.

(b) Since $M$ is an invariant mass term, it is presumably very large, 
corresponding to the scale of new physics responsible for its existence. 
In that case, $m_D << M$, and $m_1 \simeq -m_D^2/M$, $m_2 \simeq M$. This 
is the famous canonical seesaw mechanism \cite{seesaw}.  Theoretically, (b) 
is considered much more $natural$ than (a) because the former requires the 
$imposition$ of an exactly conserved global U(1) symmetry, i.e. lepton 
number. Consequently, (b) dominated the thinking on neutrino mass for many 
years until somewhat recently.

\subsection{TYPE II SEESAW} 

Another just as $natural$ way to obtain a small Majorana neutrino mass 
is to add a Higgs triplet $(\xi^{++},\xi^+,\xi^0)$ which couples directly 
to the symmetric triplet combination of two $(\nu,l)_L$ doublets, i.e.
\begin{equation}
\frac{h_\nu}{2} \left[ \nu \nu \xi^0 - \frac{(\nu l + l \nu)}{\sqrt{2}} \xi^+ 
+ l l \xi^{++} \right] \Rightarrow m_\nu = h_\nu \langle \xi^0 \rangle,
\end{equation}
with $\langle \xi^0 \rangle << \langle \phi^0 \rangle$.  It is often 
$mistakenly$ assumed that this requires extreme fine tuning and is thus not 
very natural.  To see how this mechanism really works \cite{ms98-1}, consider 
the most general Higgs potential of $\Phi$ and $\xi$:
\begin{eqnarray}
V &=& m^2 \Phi^\dagger \Phi + M^2 \xi^\dagger \xi  + \frac{1}{2} \lambda_1 
(\Phi^\dagger \Phi)^2 + \frac{1}{2} \lambda_2 (\xi^\dagger \xi)^2 + \lambda_3 
(\xi^\dagger \xi^\dagger)(\xi \xi) \nonumber \\ 
&+& f_1 (\Phi^\dagger \Phi) (\xi^\dagger \xi) + f_2 (\xi^\dagger \Phi) 
(\Phi^\dagger \xi) + [\mu \xi^\dagger \Phi \Phi + H.c.]
\end{eqnarray}
Let $\langle \phi^0 \rangle = v$, $\langle \xi^0 \rangle = u$, then
\begin{eqnarray}
v [m^2 + \lambda_1 v^2 + (f_1 + f_2) u^2 + 2 \mu u] = 0, ~~~
u [M^2 + \lambda_2 u^2 + (f_1 + f_2) v^2] +  \mu v^2 = 0.
\end{eqnarray}
If lepton number is imposed on $V$ \cite{gr81}, then $\mu=0$, and for 
both $v$ and $u$ to be nonzero, they must be given by
\begin{equation}
v^2 = \frac{-\lambda_2 m^2 + (f_1 + f_2) M^2}{\lambda_1 \lambda_2 - 
(f_1 + f_2)^2}, ~~~ 
u^2 = \frac{(f_1 + f_2) m^2 - \lambda_1 M^2}{\lambda_1 \lambda_2 - 
(f_1 + f_2)^2}.
\end{equation}
Since $u$ has to be tiny, extreme fine tuning is required.  In addition, this 
model breaks lepton number spontaneously, which implies the existence of a 
massless Goldstone particle, the majoron, i.e. $\sqrt{2}$Im$\xi^0$.  Now the 
mass of $\sqrt{2}$Re$\xi^0$ is of order $u$, hence the invisible decay $Z \to$ 
$\sqrt{2}$Re$\xi^0$ + $\sqrt{2}$Im$\xi^0$ is expected and its rate is 
equivalent to that of two neutrino pairs. This has been ruled out 
experimentally for more than 20 years.
 
For $\mu \neq 0$, a completely $new$ and $naturally$ $small$ solution for 
$u$ appears:
\begin{equation}
v^2 \simeq \frac{-m^2}{\lambda_1}, ~~~ u \simeq \frac{-\mu v^2}{M^2 + 
(f_1 + f_2) v^2}.
\end{equation}
This is now commonly called the Type II seesaw.  It works because the 
spontaneous breaking of electroweak symmetry is already accomplished by 
$\langle \phi^0 \rangle$, hence $\langle \xi^0 \rangle$ may be small, 
as long as $m^2_\xi$ is positive and large.  The parameter $\mu$ (which 
has the dimension of mass) may also be $naturally$ small, because its 
absence enhances the symmetry of $V$. In the past ten years, 
this mechanism is being appreciated more, and is now competitive 
with the Type I seesaw.

\section{SIX GENERIC MECHANISMS}

In 1979, Weinberg showed \cite{w79} that in the Minimal Standard Model, 
there is only one effective dimension-five operator, i.e.
\begin{equation}
{\cal L}_5 = \frac{f_{\alpha \beta}}{2 \Lambda} (\nu_\alpha \phi^0 - 
l_\alpha \phi^+) (\nu_\beta \phi^0 - l_\beta \phi^+),
\end{equation}
and it generates a small Majorana neutrino mass given by $f_{\alpha \beta} 
v^2/\Lambda$, where $\Lambda$ is a large effective mass.  This shows that 
all Majorana neutrino masses in the SM are necessarily $seesaw$:  for 
$v$ fixed, $m_\nu$ goes down as $\Lambda$ goes up.

In 1998, I showed \cite{m98-1} that there are three and only three ways to 
obtain the Weinberg operator at tree level, as shown in FIG.~1, and that 
there are three generic mechanisms in one-loop order.

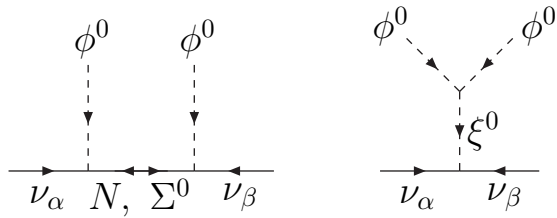
\begin{figure}[htb]
\begin{picture}(360,80)(0,0)
\ArrowLine(80,10)(110,10)
\ArrowLine(180,10)(150,10)
\ArrowLine(140,10)(110,10)
\ArrowLine(120,10)(150,10)
\DashArrowLine(110,50)(110,10)3
\DashArrowLine(150,50)(150,10)3
\ArrowLine(220,10)(250,10)
\ArrowLine(280,10)(250,10)
\DashArrowLine(250,40)(250,10)3
\DashArrowLine(230,60)(250,40)3
\DashArrowLine(270,60)(250,40)3
\Text(95,0)[]{\large$\nu_\alpha$}
\Text(130,0)[]{\large$N,~\Sigma^0$}
\Text(168,0)[]{\large$\nu_\beta$}
\Text(112,60)[]{\large$\phi^0$}
\Text(152,60)[]{\large$\phi^0$}
\Text(225,65)[]{\large$\phi^0$}
\Text(281,65)[]{\large$\phi^0$}
\Text(260,25)[]{\large$\xi^0$}
\Text(235,0)[]{\large$\nu_\alpha$}
\Text(268,0)[]{\large$\nu_\beta$}
\end{picture}
\caption{Three tree-level realizations of seesaw Majorana neutrino mass.}
\end{figure}

\subsection{Type I}

This is the canonical seesaw.  Instead of using the $2 \times 2$ matrix of 
Eq.~(3) which you learned in text books, consider the Feynman diagram of 
FIG.~1.  Just read off the neutrino mass from the two couplings of $\nu_L$ 
to $\phi^0$, each multiplied by $\langle \phi^0 \rangle$, and divided by 
the large Majorana mass of the neutral fermion singlet $N$.  The insertion 
of $N$ is obvious in the explicit structure of Eq.~(9).

\subsection{Type II}

This is obtained by coupling two lepton doublets to a scalar triplet 
$(\xi^{++},\xi^+,\xi^0)$ as shown in Eq.~(4), and linked to $\Phi \Phi$ 
as shown in Eq.~(5).  It generates the Weinberg operator as well because
\begin{equation}
(\nu_\alpha \phi^0 - l_\alpha \phi^+)(\nu_\beta \phi^0 - l_\beta \phi^+) = 
\nu_\alpha \nu_\beta \phi^0 \phi^0 - (\nu_\alpha l_\beta + l_\alpha \nu_\beta) 
\phi^+ \phi^0 + l_\alpha l_\beta \phi^+ \phi^+.
\end{equation}
Note that the decay branching fractions of $\xi^{++} \to l^+_\alpha l^+_\beta$ 
would be proportional to the entries of the neutrino mass matrix 
$({\cal M}_\nu)_{\alpha \beta}$, and may be verifiable \cite{mrs00} at 
the Large Hadron Collider (LHC).

\subsection{Type III}

Replace the singlet $N$ by the triplet $(\Sigma^+,\Sigma^0,\Sigma^-)$ 
\cite{flhj89}, then this is again obtained with the Weinberg operator because
\begin{equation}
(\nu_\alpha \phi^0 - l_\alpha \phi^+)(\nu_\beta \phi^0 - l_\beta \phi^+) = 
- 2 \nu_\alpha \phi^+ l_\beta \phi^0 + (\nu_\alpha \phi^0 + l_\alpha \phi^+) 
(\nu_\beta \phi^0 + l_\beta \phi^+) - 2 l_\alpha \phi^0 \nu_\beta \phi^+.
\end{equation}
This mechanism was largely neglected until recently.  For a recent review, 
see Ref.~\cite{m08-3}.

\subsection{Type IV, V, VI}

There are also three generic one-loop mechanisms \cite{m98-1}.  Consider a 
loop linking $\nu_L$ to $\nu_L$.  It should have an internal fermion line, 
as well as an internal scalar line.  The two external $\phi^0$ lines of the 
Weinberg operator may then be chosen to be attached in three different ways: 
one to the fermion line and one to the scalar line (Type IV), two to the 
scalar line (Type V), and two to the fermion line (Type VI).  
Almost all models of one-loop neutrino mass are of Type IV, the most 
well-known of which is the Zee model \cite{z80}.  Type V models used to be 
quite rare.  Since 2006, the idea \cite{m06-1} that neutrino mass comes from 
dark matter in one loop (scotogenic) requires precisely this mechanism.
Type VI models are unknown, presumably because they are rather complicated 
to realize.

\section{MINIMAL MODELS}

\noindent \{1\} The canonical approach is to add three neutral singlet 
fermions $N_{1,2,3}$ without imposing additive lepton number.  Majorana 
neutrino masses are then obtained via mechanism (I) and a conserved 
multiplicative $Z_2$ lepton number $(-)^L$ emerges naturally.  The most 
important consequence is the occurrence of neutrinoless double beta decay: 
$$ d \rightarrow u e^- \{ \nu_e \nu_e \} e^- u \leftarrow d.$$
In the $3 \times 3$ Majorana neutrino mass matrix in the $(e,\mu,\tau)$ 
basis, the effective neutrino mass $m_{ee}$ can be read off as the 
\{$\nu_e \nu_e$\} entry.\\

\noindent \{2\} If $N_{1,2,3}$ are added with the imposition of additive lepton 
number $L$, then $m_{ee} = 0$.  Of course, if $L$ is violated by other 
interactions, there will be a contribution to $m_{ee}$, as well as to 
neutrinoless double beta decay, but the former may not be the dominant 
cause of the latter.\\

\noindent \{3\} Instead of $N_{1,2,3}$, one Higgs scalar triplet 
$(\xi^{++},\xi^+, \xi^0)$ is added with $L=-2$.  Here $\xi^0$ couples directly 
to $\nu_L \nu_L$, and a Majorana neutrino mass is obtained if $\langle \xi^0 
\rangle \neq 0$. However, $L$ is then broken spontaneously, resulting in a 
massless Goldstone boson, i.e. the triplet majoron $\sqrt{2}$Im$\xi^0$.  At 
the same time, the scalar boson $\sqrt{2}$Re$\xi^0$ has a mass of order 
$\langle \xi^0 \rangle$ so that the decay width of $Z \to \sqrt{2}$Im$\xi^0$ + 
$\sqrt{2}$Re$\xi^0$ is equal to that of 2 neutrinos, thus ruled out 
by the well-measured invisible width of the $Z$.\\

\noindent \{4\} If $(\xi^{++},\xi^+,\xi^0)$ is added with $(-)^L = +$, then 
$\langle \xi^0 \rangle$ does not break $(-)^L$ and all neutrino masses are 
Majorana via mechanism (II).  If $M_\xi$ is of order 1 TeV, the decay 
$\xi^{++} \to l^+_i l^+_j$ will map out the relative magnitudes of all 
elements of the neutrino mass matrix \cite{mrs00}.\\

\noindent \{5\} Instead of $N$, Majorana triplet fermions $(\Sigma^+,
\Sigma^0,\Sigma^-)$ are added.  Neutrino masses are then obtained via 
mechanism (III). Mixing occurs between $l$ and $\Sigma$ as well.\\

\noindent \{6\} If just one $N$ is added, then a particular linear combination 
of $\nu_{e,\mu,\tau}$, call it $\nu_1$, will couple to $N$ and gets a seesaw 
mass.  The other two linear combinations are massless at tree level, but 
since there is no symmetry which prevents it, they will become massive 
in two loops \cite{bm88}, to be discussed in the next section.\\

\noindent \{7\} Consider \cite{bgl89} the addition of one $N$ and a second 
scalar doublet $(\eta^+,\eta^0)$ without any symmetry restriction.  Define 
$\Phi$ to be the scalar doublet with vacuum expectation value and $\eta$ 
the one without, then the Yukawa terms $f_{11} \bar{\nu}_{1L} N_R \bar{\phi}^0 
+ f_{12} \bar{\nu}_{1L} N_R \bar{\eta}^0 + f_{22} \bar{\nu}_{2L} N_R 
\bar{\eta}^0$ imply $\nu_1$ gets a tree-level mass, $\nu_2$ gets a radiative 
mass via mechanism (V), and $\nu_3$ gets a two-loop mass mentioned in \{6\}.\\

\noindent \{8\} Consider \cite{m01} the addition of $N_{1,2,3}$ with $L=0$ and 
$(\eta^+,\eta^0)$ with $L=-1$. Then $(\nu \phi^0 - l \phi^+)N$ is forbidden 
but $(\nu \eta^0 - l \eta^+)N$ is allowed.  Let $L$ be broken softly in 
the scalar sector by the unique term $\mu^2 (\Phi^\dagger \eta + \eta^\dagger 
\Phi)$.  The Higgs potential is then given by
\begin{eqnarray}
V &=& m_1^2 \Phi^\dagger \Phi + m_2^2 \eta^\dagger \eta + \frac{1}{2} \lambda_1 
(\Phi^\dagger \Phi)^2 + \frac{1}{2} \lambda_2 (\eta^\dagger \eta)^2 \nonumber 
\\ &+& \lambda_3 (\Phi^\dagger \Phi)(\eta^\dagger \eta) + \lambda_4 
(\Phi^\dagger \eta)(\eta^\dagger \Phi) + \mu^2 (\Phi^\dagger \eta + 
\eta^\dagger \Phi).
\end{eqnarray}
Let $\langle \phi^0 \rangle = v$ and $\langle \eta^0 \rangle = u$, then
\begin{eqnarray}
v [m_1^2 + \lambda_1 v^2 + (\lambda_3 + \lambda_4) u^2] + \mu^2 u = 0, ~~~
u [m_2^2 + \lambda_2 u^2 + (\lambda_3 + \lambda_4) v^2] + \mu^2 v = 0.
\end{eqnarray}
For $m_1^2 < 0$ and $m_2^2 > 0$ and large, the solution is
\begin{equation}
v^2 \simeq \frac{-m_1^2}{\lambda_1}, ~~~ u \simeq \frac{-\mu^2 v}{m_2^2 + 
(\lambda_3 + \lambda_4) v^2}.
\end{equation}
The Majorana neutrino mass is still given by the seesaw formula 
$m_\nu \simeq -m_D^2/m_N$, but $m_D$ is now proportional to $u$ which is 
small because $\mu^2$ is a $L$ violating parameter and can be chosen 
to be small naturally.  For example, if $u \sim 1$ MeV, $m_N \sim 1$ TeV, 
then $u^2/m_N \sim 1$ eV is the neutrino mass scale.

\section{RADIATIVE MODELS}

Neutrino masses may be generated in one loop or more, depending on the 
assumed particle content beyond the minimal SM, and additional possible 
symmetries. Here I discuss four examples, three early and one recent.

\subsection{Generic 2-W Mechanism in the SM}
\begin{figure}[htb]
\begin{center}
\begin{picture}(400,120)(0,0)
\ArrowLine(80,60)(120,60)
\ArrowLine(120,60)(150,60)
\ArrowLine(150,60)(180,60)
\ArrowLine(210,60)(180,60)
\ArrowLine(240,60)(210,60)
\ArrowLine(280,60)(240,60)

\PhotonArc(165,60)(45,0,180)39
\PhotonArc(195,60)(45,180,0)39

\Text(100,50)[]{\large$\nu_{2,3}$}
\Text(260,50)[]{\large$\nu_{2,3}$}
\Text(135,50)[]{\large$l$}
\Text(225,50)[]{\large$l$}
\Text(165,50)[]{\large$\nu_1$}
\Text(195,50)[]{\large$\nu_1$}
\Text(165,115)[]{\large$W$}
\Text(195,5)[]{\large$W$}

\end{picture}
\end{center}
\caption{Two-$W$ generation of neutrino mass.}
\end{figure}
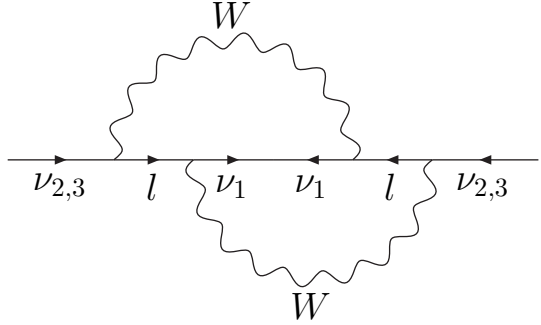
The minimal model for all neutrinos to acquire mass is to add just one 
$N$ as discussed in \{6\} of the previous section.  In that case, only 
one linear combination of $\nu_{e,\mu.\tau}$, call it $\nu_1$, gets a 
tree-level Majorana mass.  The others appear to be massless, but that 
cannot be so, because if $\nu_1$ spans all three flavors, there is no 
remaining symmetry which can keep them massless.  On the other hand, only 
SM particles are available, so how in the world can they acquire mass?  
The answer was provided by Ref.~\cite{bm88} where it was shown that these 
masses appear in two loops, from the exchange of two $W$ bosons, as shown 
in FIG.~2. This diagram is doubly suppressed by the Glashow-Ilioupoulos-Maiani 
mechanism \cite{gim70} and yields extremely small neutrino masses.

\subsection{Zee/Wolfenstein Model}

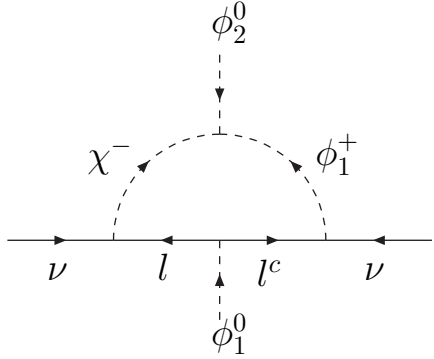
\begin{figure}[htb]
\begin{center}\begin{picture}(500,100)(120,25)
\ArrowLine(270,50)(310,50)
\ArrowLine(350,50)(390,50)
\ArrowLine(350,50)(310,50)
\ArrowLine(430,50)(390,50)
\Text(290,35)[b]{\large$\nu$}
\Text(410,35)[b]{\large$\nu$}
\Text(330,35)[b]{\large$l$}
\Text(370,33)[b]{\large$l^c$}
\Text(310,75)[b]{\large$\chi^-$}
\Text(396,75)[b]{\large$\phi_1^+$}
\DashArrowArc(350,50)(40,0,90){3}
\DashArrowArcn(350,50)(40,180,90){3}
\DashArrowLine(350,20)(350,50){3}
\Text(355,7)[b]{\large$\phi_1^0$}
\DashArrowLine(350,120)(350,90){3}
\Text(355,125)[b]{\large$\phi_2^0$}
\end{picture}
\end{center}
\caption[]{One-loop radiative neutrino mass.}
\end{figure}
The most well-known radiative model \cite{z80} uses the fact that the 
invariant combination of two (different) lepton doublets couples to a 
charged scalar singlet, i.e. $(\nu_i l_j - l_i \nu_j) \chi^+$.  By the same 
token, two different scalar doublets are also needed, i.e. $(\phi_1^+ \phi_2^0 
- \phi_1^0 \phi_2^+) \chi^-$.  In that case, radiative neutrino masses are 
obtained in one loop via mechanism (IV), where $\Phi_2$ has been assumed not 
to couple to leptons \cite{w80}, as shown in FIG.~3.  The $3 \times 3$ 
neutrino mass matrix is then of the form
\begin{equation}
{\cal M}_\mu = \begin{pmatrix}0 & f_{\mu e} (m_\mu^2 - m_e^2) & f_{\tau e} 
(m_\tau^2 - m_e^2) \\  f_{\mu e} (m_\mu^2 - m_e^2) & 0 & f_{\tau \mu} (m_\tau^2 
- m_\mu^2) \\  f_{\tau e} (m_\tau^2 - m_e^2) & f_{\tau \mu} (m_\tau^2 - m_\mu^2) 
& 0 \end{pmatrix}
\end{equation}
This model was studied intensively, but it is now ruled out by data.
  
\subsection{Zee/Babu Model}

\begin{figure}[htb]
\begin{center}
\begin{picture}(350,80)(20,30)
\ArrowLine(80,60)(120,60)
\ArrowLine(150,60)(120,60)
\ArrowLine(150,60)(180,60)
\ArrowLine(210,60)(180,60)
\ArrowLine(210,60)(240,60)
\ArrowLine(280,60)(240,60)
\DashArrowLine(150,35)(150,60)5
\DashArrowLine(210,35)(210,60)5
\DashArrowLine(180,60)(180,120)5
\DashArrowArc(180,60)(60,90,180)5
\DashArrowArcn(180,60)(60,90,0)5

\Text(100,50)[]{\large$\nu$}
\Text(260,50)[]{\large$\nu$}
\Text(135,50)[]{\large$l$}
\Text(165,50)[]{\large$l^c$}
\Text(195,50)[]{\large$l^c$}
\Text(225,50)[]{\large$l$}
\Text(150,25)[]{\large$\phi^0$}
\Text(210,25)[]{\large$\phi^0$}
\Text(235,110)[]{\large$\chi^+$}
\Text(132,110)[]{\large$\chi^+$}
\Text(195,90)[]{\large$\zeta^{++}$}

\end{picture}
\end{center}
\caption{Two-loop radiative neutrino mass.}
\end{figure}
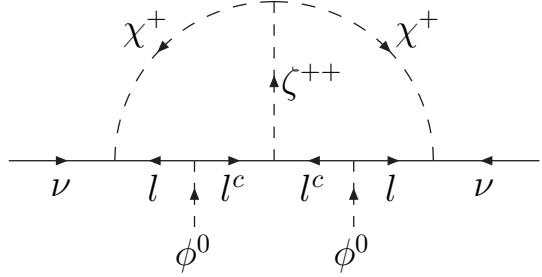
In the previous model, if the second Higgs doublet is replaced by a doubly 
charged singlet $\zeta^{++}$, a two-loop neutrino mass is obtained 
\cite{z86,b88}, using the additional interactions $\zeta^{++} \chi^- \chi^-$ 
and $l^c_i l^c_j \zeta^{--}$.  Note that it is doubly suppressed by lepton 
masses as in Eq.~(15).  However, nonzero diagonal entries 
are now allowed in the neutrino mass matrix and there are enough free 
parameters not to be ruled out.  Processes such as $\mu \to eee$ and 
$\tau \to \mu \mu \mu, \mu \mu e, \mu e e, e e e$ are possible at tree 
level and act as constraints as well as opportunities for discoveries.

\subsection{Scotogenic Neutrino Mass}

A recent new development \cite{m06-1,m07-1} is to connect the origin of 
neutrino mass to the existence of dark matter, i.e. scotogenic.  The idea 
is very simple. Let the SM be extended to include three $N$'s and a second 
scalar doublet $(\eta^+,\eta^0)$ \cite{dm78} which are odd under a new 
exactly conserved $Z_2$ discrete symmetry, whereas all SM particles are even. 
In that case, the usual Yukawa term $(\nu \phi^0 - l \phi^+)N$ is forbidden, 
but $(\nu \eta^0 - l \eta^+)N$ is allowed.  However, unlike the case \{8\} 
discussed in Section 3, $\langle \eta^0 \rangle = 0$ here because of the 
conserved $Z_2$.  Hence there is no $m_D$ linking $\nu$ and $N$.  However, 
$\nu$ gets a radiative Majorana mass (of Type V) directly as shown in FIG.~5.
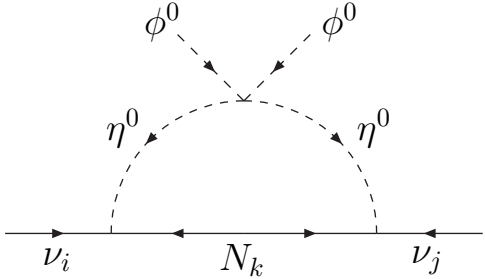
\begin{figure}[htb]
\begin{center}
\begin{picture}(360,120)(0,0)
\ArrowLine(90,10)(130,10)
\ArrowLine(180,10)(130,10)
\ArrowLine(180,10)(230,10)
\ArrowLine(270,10)(230,10)
\DashArrowLine(155,85)(180,60)3
\DashArrowLine(205,85)(180,60)3
\DashArrowArc(180,10)(50,90,180)3
\DashArrowArcn(180,10)(50,90,0)3

\Text(110,0)[]{\large$\nu_i$}
\Text(250,0)[]{\large$\nu_j$}
\Text(180,0)[]{\large$N_k$}
\Text(135,50)[]{\large$\eta^0$}
\Text(230,50)[]{\large$\eta^0$}
\Text(150,90)[]{\large$\phi^{0}$}
\Text(217,90)[]{\large$\phi^{0}$}

\end{picture}
\end{center}
\caption{One-loop scotogenic neutrino mass.}
\end{figure}

\noindent Specifically, this diagram is exactly calculable from the 
exchange of $\sqrt{2}$Re$\eta^0$ and $\sqrt{2}$Im$\eta^0$, i.e.
\begin{equation}
({\cal M}_\nu)_{ij} = \sum_k \frac{h_{ik} h_{jk} M_k}{16 \pi^2} \left[ 
\frac{m_R^2}{m_R^2-M_k^2} \ln \frac{m_R^2}{M_k^2} - \frac{m_I^2}{m_I^2-M_k^2} 
\ln \frac{m_I^2}{M_k^2} \right].
\end{equation}
If $\sqrt{2}$Re$\eta^0$ or $\sqrt{2}$Im$\eta^0$ is the lightest particle of 
odd $Z_2$, then it is a possible dark-matter candidate 
\cite{m06-1,bhr06,lnot07,glbe07} and may be searched for at the LHC 
\cite{cmr07}.

\section{GENERIC CONSEQUENCES OF NEUTRINO MASS}

\noindent (A) Once neutrinos have mass and mix with one another, the radiative 
decay $\nu_2 \to \nu_1 \gamma$ happens in all models, but is usually harmless 
as long as $m_\nu < $ few eV, in which case it will have an extremely long 
lifetime, many orders of magnitude greater than the age of the Universe.

\noindent (B) The analogous radiative decay $\mu \to e \gamma$ also happens 
in all models, but is only a constraint for some models where $m_\nu$ is 
radiative in origin.

\noindent (C) Neutrinoless double beta decay occurs, proportional to the 
\{$\nu_e \nu_e$\} entry of the Majorana meutrino mass matrix.

\noindent (D) Leptogenesis is possible from $N \to l^+ \phi^- (l^- \phi^+)$ 
or $\xi^{++} \to l^+ l^+ (\phi^+ \phi^+)$.  There may also be other 
possibilities.

\noindent (E) New particles at the 100 GeV mass scale exists in some models. 
They can be searched for at the LHC and beyond.

\noindent (F) Lepton-flavor changing processes at tree level may provide 
subdominant contributions to neutrino oscillations.

\noindent (G) Lepton-number violating interactions at the TeV mass scale may 
erase any pre-existing $B$ or $L$ asymmetry of the Universe.

\section{LEPTON NUMBER IN SUPERSYMMETRY}

In the SM (without $\nu_R$) , neutrinos are massless and four global U(1) 
symmetries are automatically conserved: $B$, $L_e$, $L_\mu$, $L_\tau$.  In 
its supersymmetric extension, the lepton doublet superfields $L_i = 
(\nu_i,l_i)$ transform exactly like one of the two Higgs superfields, i.e. 
$\Phi_1 = (\phi_1^0,\phi_1^-)$.  To tell them apart, lepton number has to be 
imposed, so that $L_i$ have $L=1$, $l^c_i$ have $L=-1$, and $\Phi_{1,2}$ have 
$L=0$.  Again, neutrinos are massless and $B$, $L_e$, $L_\mu$, $L_\tau$ 
are conserved.  Thus $R$ parity, defined as $(-)^{3B+L+2j}$, is also 
conserved and the model is known as the Minimal Supersymmetric Standard 
Model (MSSM).

The terms in the superpotential which conserve $B$ but not $L$ may be 
organized to allow for 5 generic scenarios \cite{mn90}. Let
\begin{eqnarray}
W^{(1)} &=& h_i \Phi_1 L_i l^c_i + h^u_{ij} \Phi_2 Q_i u^c_j + h^d_{ij} \Phi_1 
Q_i d^c_j + \mu_0 \Phi_1 \Phi_2, \\ 
W^{(2)} &=& f_e L_3 L_1 l^c_1 + f_\mu L_3 L_2 l^c_2 + f_{ij} L_3 Q_i d^c_j 
+ \mu_3 L_3 \Phi_2, \\ 
W^{(3)} &=& f_{e \mu \tau} L_1 L_2 l^c_3, \\ 
W^{(4)} &=&  f_{e \mu} L_3 L_1 l^c_2 + f_{\mu e} L_3 L_2 l^c_1, \\ 
W^{(5)} &=& f'_e L_2 L_1 l^c_1 + f'_\tau L_2 L_3 l^c_3 + f'_{ij} L_2 Q_i d^c_j 
+ \mu_2 L_2 \Phi_2,
\end{eqnarray}
then the following 5 models are possible (each with 3 obvious permutations):

\noindent (A) $W^{(1)} + W^{(2)} \Rightarrow L_e, L_\mu$ conserved, but 
$\nu_\tau$ mixes with the other 4 neutralinos, and gets a seesaw mass:
\begin{equation}
m_{\nu_\tau} = \frac{\mu_3^2}{2 \mu_0 \tan \beta} \left[ 1 - 
\frac{\mu_0 M_1 M_2} {M_Z^2 (c^2 M_1 + s^2 M_2) \sin 2 \beta} \right]^{-1}.
\end{equation}

\noindent (B) $W^{(1)} + W^{(3)} \Rightarrow L_e, L_\mu, L_\tau$ conserved, 
with $L_\tau = L_e + L_\mu$.  This is the simplest new model with just one 
term in $W^{(3)}$.  Neutrinos remain massless.  The $\tilde{W}^+$ gaugino 
will decay into $e^+ \mu^+ \tau^-$ via $\tilde{\nu}_e$ and $\tilde{\nu}_\mu$.

\noindent (C) $W^{(1)} + W^{(2)} + W^{(3)} \Rightarrow L_e, L_\mu (L_\mu = 
-L_e)$ conserved.

\noindent (D) $W^{(1)} + W^{(2)} + W^{(4)} \Rightarrow L_e, L_\mu (L_\mu = 
L_e)$ conserved.

\noindent (E) $W^{(1)} + W^{(2)} + W^{(5)} \Rightarrow L_e$ conserved only. 
One neutrino gets a tree-level mass as in (A),(C),(D), another gets a 
radiative mass.

\section{U(1) GAUGE SYMMETRIES}

The SM may be extended to include an extra $U(1)_X$ gauge symmetry.  This 
requires the absence of quantum anomalies:

\noindent (A) Mixed gravitational-gauge anomaly: The sum of $U(1)_X$ charges 
should be zero.

\noindent (B) Global SU(2) anomaly: The number of SU(2) fermion doublets 
should be even.

\noindent (C) Axial-vector-vector-vector anomaly:  The sum over $L^3 - R^3$ 
charges, i.e. $[SU(3)_C]^2 U(1)_X$, $[SU(2)_L]^2 U(1)_X$, $[U(1)_Y]^2 U(1)_X$, 
$U(1)_Y [U(1)_X]^2$, and $[U(1)_X]^3$, should be zero.

Given the particle content of the SM (without $\nu_R$), there are 3 often 
neglected possible $U(1)_X$ gauge extensions \cite{hjlv91}:  $L_e-L_\mu$, 
$L_e-L_\tau$, $L_\mu-L_\tau$.  If 3 $\nu_R$'s are added with $L=1$, then 
$U(1)_X = B - L$ is possible.  The anomaly conditions are satisfied as follows:
\begin{eqnarray}
X &:& (3)(2)\left[ \frac{1}{3} - \frac{1}{3} \right] + (2) [(-1) - (-1)] 
= 0. \\ ~
C^2 X &:& \frac{1}{2} (2) \left[ \frac{1}{3} - \frac{1}{3} \right] 
= 0. \\ ~ 
L^2 X &:& \frac{1}{2} \left[ (3) \left( \frac{1}{3} \right) + (-1) 
\right] = 0. \\ ~
Y^2 X &:& \left[ 2 \left( \frac{1}{6} \right)^2 - \left( 
\frac{2}{3} \right)^2 - \left( -\frac{1}{3} \right)^2 \right] \left( 
\frac{1}{3} \right) + \left[ 2 \left( -\frac{1}{2} \right)^2 - (-1)^2 
\right] (-1) = 0. \\ ~
Y X^2 &:& (3) \left[ 2 \left( \frac{1}{6} \right) - \left( 
\frac{2}{3} \right) - \left( -\frac{1}{3} \right) \right] \left( \frac{1}{3} 
\right)^2 + \left[ 2 \left( -\frac{1}{2} \right) - (-1) \right] (-1)^2 = 0. \\~
X^3 &:& (3) (2) \left[ \left( \frac{1}{3} \right)^3 - \left( 
\frac{1}{3} \right)^3 \right] + (2) [(-1)^3 - (-1)^3] = 0.
\end{eqnarray}
Neutrino mass may thus be a hint of $U(1)_{B-L}$ and point to $SU(3)_C \times 
SU(2)_L \times SU(2)_R \times U(1)_{B-L}$ and $SO(10)$.

\section{$B-3L_\tau$}

If just one $\nu_R$ with $L_\tau = 1$ is added, then $U(1)_X = B - 3L_\tau$ is 
anomaly-free and can be gauged \cite{m98-2,mr98}.  To break $U(1)_X$ 
spontaneously, a neutral scalar singlet $\chi^0 \sim (1,1,0;6)$ is used, which 
also gives $\nu_R$ a large Majorana mass, thereby making $\nu_\tau$ massive. 
The $X$ boson decays into quarks and $\tau$ but not $e$ or $\mu$.  If a 
scalar doublet $(\eta^+,\eta^0) \sim (1,2,1/3;-3)$ and a scalar singlet 
$\chi^- \sim (1,1,-1;-3)$ are added, then one linear combination of 
$\nu_e,\nu_\mu,\nu_\tau$ gets a tree-level mass, and the others get 
radiative masses via the radiative mechanism of Type IV.

The $X$ boson is not constrained to be very heavy because it does not couple 
to $e$ or $\mu$.  It can be produced easily at the LHC because it has quark 
couplings.  Its decay into $\tau^+ \tau^-$ is also a good signature. 
Realistic neutrino masses and mixing are possible, with additional $U(1)_X$ 
scalars.  It is well-known that $B-L$ may come from $SU(4) \times SU(2)_L 
\times SU(2)_R$ with $Q = T_{3L} + T_{3R} + (B-L)/2$ and $SU(4)$ breaking to 
$SU(3)_C \times U(1)_{B-L}$.  Analogously, $B-3L_\tau$ may come from 
$SU(10) \times SU(2)_L \times U(1)_{Y'}$ with $Q = T_{3L} + Y' + (B-3L_\tau)/5$ 
and $SU(10)$ breaking to $[SU(3)_C]^3 \times U(1)_{B-3L_\tau}$.

\section{$U(1)_\Sigma$}

Instead of using the Type I seesaw for neutrino mass, consider Type III by 
adding 3 copies of the fermion triplet $(\Sigma^+,\Sigma^0,\Sigma^-)_R \sim 
(1,3,0)$.  Is there a U(1) gauge symmetry like $B-L$ as in the case of 
$\nu_R$?  The answer is yes \cite{m02-1,mr02,bd05,aem09}.  Call this 
$U(1)_\Sigma$ and let $(u,d)_L \sim n_1$, $u_R \sim n_2$, $d_R \sim n_3$, 
$(\nu,e)_L \sim n_4$, $e_R \sim n_5$, and $\Sigma_R \sim n_6$.  Then the 
6 conditions for $U(1)_\Sigma$ to be anomaly-free, including the highly 
nontrivial 
\begin{equation}
6n_1^3 - 3n_2^3 - 3n_3^3 + 2n_4^3 - n_5^3 - 3n_6^3 = 0,
\end{equation}
are satisfied with
\begin{equation}
4n_2 = 7n_1 - 3n_4, ~~~ 4n_3 = n_1 + 3n_4, ~~~ 4n_5 = -9n_1 + 5n_4, ~~~ 
4n_6 = 3n_1 + n_4.
\end{equation}
This is a very remarkable result.

There is thus a family of solutions defined by $n_4 = \lambda n_1$. If 
$\lambda = -3$ and $n_1$ is chosen to be 1/6 for convenience, then 
$U(1)_\Sigma = U(1)_Y$, but if $\lambda \neq -3$, then $U(1)_\Sigma$ is new. 
Two Higgs doublets are required for fermion masses: $(\phi_1^+,\phi_1^0) 
\sim n_1-n_3 = n_2-n_1 = n_6-n_4 = 3(n_1-n_4)/4$ couples to quarks and 
$\Sigma$, and $(\phi_2^+,\phi_2^0) \sim n_4-n_5 = (9n_1-n_4)/4$ couples to $e$. 
Since $2 \Sigma_R^+ \Sigma_R^- + \Sigma_R^0 \Sigma_R^0$ is an SM invariant, 
$\Sigma_R$ may obtain a large Majorana mass just as $\nu_R$.  It mixes 
necessarily  with $(\nu,e)_L$ through $\Phi_1$, so that $\Sigma_i^- \to 
e_j^- Z$ and $\nu_j W^-$ are possible signals at the LHC.

\section{SUPERSYMMETRIC $U(1)_X$}

If the SM is extended to include supersymmetry, 3 well-known issues spring up. 
(A) $m_\nu = 0$ as in the SM. (B) $B$ and $L_i$ are conserved only if imposed. 
(C) The allowed term $\mu \hat{\phi}_1 \hat{\phi}_2$ in the superpotential 
must be adjusted with $\mu \sim M_{SUSY}$, i.e. the supersymmetry breaking 
scale. Each has a piecemeal solution, but is there one unifying explanation 
using $U(1)_X$?  The answer is again yes \cite{m02-2}. Here all superfields 
must be considered in the anomaly-free conditions.  Under $U(1)_X$, let there 
be 3 copies of
\begin{eqnarray}
&& (\hat{u},\hat{d}) \sim (3,2,\frac{1}{6};n_1), ~~~ \hat{u}^c \sim 
(3^*,1,-\frac{2}{3};n_2), ~~~ \hat{d}^c \sim (3^*,1,\frac{1}{3};n_3), \\ 
&& (\hat{\nu},\hat{e}) \sim (1,2,-\frac{1}{2};n_4), ~~~ \hat{e}^c \sim 
(1,1,1;n_5), ~~~ \hat{N}^c \sim (1,1,0;n_6),
\end{eqnarray}
and 1 copy of
\begin{equation}
\hat{\phi}_1 \sim (1,2,-\frac{1}{2};-n_1-n_3), ~~~ \hat{\phi}_2 \sim 
(1,2,\frac{1}{2};-n_1-n_2),
\end{equation}
with $n_1+n_3=n_4+n_5$ and $n_1+n_2=n_4+n_6$, so that quarks and leptons 
obtain masses through the two scalar superfields as in the MSSM.
The Higgs singlet superfield
\begin{equation}
\hat{\chi} \sim (1,1,0;2n_1+n_2+n_3)
\end{equation}
is then added, so that $\mu \hat{\phi}_1 \hat{\phi}_2$ is replaced by 
$\hat{\chi} \hat{\phi}_1 \hat{\phi}_2$ and $\langle \chi \rangle \neq 0$ 
breaks $U(1)_X$.  $Two$ copies of singlet $up$ quark superfields
\begin{equation}
\hat{U} \sim (3,1,\frac{2}{3};n_7), ~~~ \hat{U}^c \sim (3^*,1,-\frac{2}{3};n_8),
\end{equation}
and $one$ copy of singlet $down$ quark superfields
\begin{equation} 
\hat{D} \sim (3,1,-\frac{1}{3};n_7), ~~~ \hat{D}^c \sim (3^*,1,\frac{1}{3};n_8),
\end{equation}
are added with $n_7+n_8 = -2n_1-n_2-n_3$ so that $\hat{\chi}\hat{U}\hat{U}^c$ 
and $\hat{\chi}\hat{D}\hat{D}^c$ are allowed, with $M_{U,D}$ appearing also at 
the $U(1)_X$ breaking scale. So far there are 8 numbers and 3 constraints, 
resulting in 5 independent numbers.  Consider first
\begin{equation}
[SU(3)]^2 U(1)_X ~:~ 2n_1 + n_2 + n_3 + n_7 + n_8 = 0.
\end{equation}
This is already satisfied. Consider then $[SU(2)]^2 U(1)_X$ and 
$[U(1)_Y]^2 U(1)_X$ respectively:
\begin{eqnarray}
 && 3(3n_1+n_4) + (-n_1-n_3) + (-n_1-n_3) = 7n_1-n_2-n_3+3n_4 
= 0, \\   && -n_1 + 7n_2 + n_3 + 3n_4 + 6n_5 + 6n_7 + 6n_8 
= -7n_1 + n_2 + n_3 - 3n_4 = 0.
\end{eqnarray}
These two conditions are identical, resulting in the elimination of one 
number.  Using $n_1,n_2,n_4,n_7$ as independent, consider $U(1)_Y [U(1)_X]^2$:
\begin{equation}
3n_1^2 - 6n_2^2 + 3n_3^2 - 3n_4^2 + 3n_5^2 + 3n_7^2 - 
3n_8^2 - (n_1+n_3)^2 + (n_1+n_2)^2 = 6(3n_1+n_4)(2n_1-4n_2-3n_7) = 0,
\end{equation}
which factors exactly and has two solutions.  If $3n_1 + n_4 = 0$, $U(1)_X 
= U(1)_Y$ as expected, so the condition $2n_1 - 4n_2 - 3n_7$ is chosen from 
now on.  Using $n_1,n_4,n_6$ as independent, the other 5 numbers are
\begin{eqnarray}
&& n_2 = -n_1+n_4+n_6, ~~ n_3=8n_1+2n_4-n_6, ~~ n_5=9n_1+n_4-n_6, \nonumber \\ 
&& n_7=2n_1-\frac{4}{3}n_4 - \frac{4}{3}n_6, ~~ n_8 = -11n_1 - \frac{5}{3}n_4 
+ \frac{4}{3}n_6.
\end{eqnarray}
The most nontrivial condition is
\begin{eqnarray}
[U(1)_X]^3 &:& 3[6n_1^3 + 3n_2^3 + 3n_3^3 + 2n_4^3 + n_5^3 + n_6^3] + 
3(3n_7^3 + 3n_8^3) \nonumber \\ &+& 2(-n_1-n_3)^3 + 2(-n_1-n_2)^3 + 
(2n_1+n_2+n_3)^3 \nonumber 
\\ &=& -36(3n_1+n_4)(9n_1+n_4-2n_6)(6n_1-n_4-n_6) = 0.
\end{eqnarray}
The sum of 11 cubic terms has been factorized exactly!  Two possible solutions 
are
\begin{equation}
{\rm (A)}~n_6 = \frac{1}{2}(9n_1 + n_4), ~~~ {\rm (B)}~n_6 = 6n_1-n_4.
\end{equation}
To obtain $L$ conservation automatically, the solutions are (A) $9n_1+5n_4 
\neq 0$, or (B) $3n_1+4n_4 \neq 0$.  To obtain $B$ conservation automatically, 
the conditions are (A) $7n_1+3n_4 \neq 0$, or (B) $3n_1+2n_4 \neq 0$.  If 
(A)=(B), then
\begin{equation}
n_1 = n_4 = 1, ~~~ n_2=n_3=n_5=n_6=5, ~~~ n_7=n_8=-6,
\end{equation}
and $U(1)_X$ is orthogonal to $U(1)_Y$.  However, there is still the  
mixed gravitational-gauge anomaly, i.e. the sum of $U(1)_X$ charges 
$= 6(3n_1+n_4) \neq 0$.  To cancel this without affecting the other 
conditions, add singlet superfields with charge in units of $(3n_1+n_4)$: 
one with charge 3, three ($\hat{S}^c$) with charge $-2$, and three ($\hat{N}$) 
with charge $-1$, so that $3 + 3(-2) + 3(-1) = -6$ and $27 + 3(-8) + 3(-1) 
= 0$.  Consider now the neutrino mass.  Since $L$ is conserved, this mass 
is Dirac, coming from the pairing of $\nu$ with $N^c$.  However, if 
$n_6 = 3n_1+n_4$, then the singlets ${S}^c$ and ${N}$ are exactly right to 
allow the neutrinos to acquire small seesaw Dirac masses.  In the basis
$(\nu,S^c,N,N^c)$, the $12 \times 12$ neutrino mass matrix is
\begin{equation}
{\cal M}_\mu = \begin{pmatrix}0 & 0 & 0 & m_1 \\0 & 0 & m_2 & 0 \\
0 & m_2 & 0 & M\\m_1 & 0 & M& 0\end{pmatrix},
\end{equation}
with $m_\nu = -m_1 m_2/M$.  Since $m_1$ comes from electroweak symmetry 
breaking and $m_2$ from $U(1)_X$ breaking, and $M$ is an invariant mass, 
this is a natural explanation of the smallness of $m_\nu$ just as in 
the seesaw Majorana case.

\section{Neutrino Tribimaximal Mixing}	

From neutrino-oscillation data in the past decade, it is now established 
that the neutrino mixing matrix $U_{l \nu}$ takes a particular form which is 
approximately tribimaximal.  Here I show how it can be understood in terms 
of an underlying non-Abelian discrete symmetry $A_4$.  In 1978, soon after 
the putative discovery of the third family of leptons and quarks, it was 
conjectured by Cabibbo \cite{c78} and Wolfenstein \cite{w78} independently that
\begin{equation}
U_{l \nu}^{CW} = \frac{1}{\sqrt{3}} \left( \begin{array}{ccc} 1 & 1 & 1 \\ 
1 & \omega & \omega^2 \\  1 & \omega^2 & \omega \end{array} \right),
\end{equation}
where $\omega = \exp(2 \pi i/3) = -1/2 + i \sqrt{3}/2$. This should dispel 
the ${myth}$ that everybody expected small mixing angles in the 
lepton sector as in the quark sector.  In 2002, after much neutrino 
oscillation data have been established, Harrison, Perkins, and Scott 
\cite{hps02} proposed the tribimaximal mixing matrix, i.e.
\begin{equation}
U_{l \nu}^{HPS} = \left( \begin{array}{ccc} \sqrt{2/3} & 1/\sqrt{3} & 0 \\ 
-1/\sqrt{6} & 1/\sqrt{3} & -1/\sqrt{2} \\ -1/\sqrt{6} & 1/\sqrt{3} & 
1/\sqrt{2} \end{array} \right) \sim (\eta_8, \eta_1, \pi^0),
\end{equation}
where the 3 columns are reminiscent of the meson nonet.
In 2004, I discovered \cite{m04} the simple connection:
\begin{equation}
U_{l \nu}^{HPS} = (U_{l \nu}^{CW})^\dagger \left( \begin{array}{ccc} 
1 & 0 & 0 \\ 0 & 1/\sqrt{2} & -1/\sqrt{2} \\ 0 & 1/\sqrt{2} & 1/\sqrt{2} 
\end{array} \right) \left( \begin{array}{ccc} 0 & 1 & 0 \\ 1 & 0 & 0 \\ 
0 & 0 & i \end{array} \right).
\end{equation}
This means that if
\begin{equation}
{\cal M}_l = U_{l \nu}^{CW} \left( \begin{array}{ccc} m_e & 0 & 0 \\ 
0 & m_\mu & 0 \\ 0 & 0 & m_\tau \end{array} \right) (U_R^l)^\dagger
\end{equation}
and ${\cal M}_\nu$ has $2-3$ reflection symmetry, with zero $1-2$ and $1-3$ 
mixing, i.e
\begin{equation}
{\cal M}_\nu = \left( \begin{array}{ccc} a+2b & 0 & 0 \\ 0 & a-b & d \\ 
0 & d & a-b \end{array} \right),
\end{equation}
$U_{l \nu}^{HPS}$ will be obtained, but how? Tribimaximal mixing 
means that
\begin{equation}
\theta_{13}=0, ~~~ \sin^2 2 \theta_{23} = 1, ~~~ \tan^2 \theta_{12} = 1/2.
\end{equation}
In 2002 (when HPS proposed it), world data were not precise enough to test 
this idea.  In 2004 (when I derived it), SNO data implied $\tan^2 \theta_{12} 
= 0.40 \pm 0.05$, which was not so encouraging.  Then in 2005, revised SNO 
data obtained $\tan^2 \theta_{12} = 0.45 \pm 0.05$, and tribimaximal mixing 
became a household word, unleashing a glut of papers.

\section{Tetrahedral Symmetry A$_4$}

For 3 families, one should look for a group with a \underline{3} 
representation, the simplest of which is A$_4$, the group of the even 
permutation of 4 objects.  It has 12 elements, divided into 4 equivalence 
classes, and 4 irreducible representations: \underline{1}, \underline{1}$'$, 
\underline{1}$''$, and \underline{3}, with the multiplication rule
\begin{eqnarray}
\underline{3} \times \underline{3} &=& \underline{1}~(11+22+33) + 
\underline{1}'~(11+\omega^2 22 + \omega 33) + \underline{1}''~(11 + 
\omega 22 + \omega^2 33) \nonumber \\ &+& \underline{3}~(23,31,12) 
+\underline{3}~(32,13,21).
\end{eqnarray}
A$_4$ is also the symmetry group of the regular tetrahedron, one of the 5 
perfect geometric solids in 3 dimensions and identified by Plato as ``fire'' 
\cite{m02-3}.  It is a subgroup of both SO(3) and SU(3).  The latter also 
has 2 sequences of finite subgroups which are of interest: $\Delta(3n^2)$ 
has $\Delta(12) \equiv$ A$_4$ and $\Delta(27)$; $\Delta(3n^2-3)$ has 
$\Delta(24) \equiv$ S$_4$.

There are two ways to achieve Eq.~(49).  The original proposal 
\cite{mr01,bmv03} is to assign $(\nu_i,l_i) \sim \underline{3}, ~l^c_i 
\sim \underline{1}, \underline{1}', \underline{1}''$, then with 
$(\phi^0_i,\phi^-_i) \sim \underline{3}$,
\begin{equation}
{\cal M}_l = \left( \begin{array}{ccc} h_1 v_1 & h_2 v_1 & h_3 v_1 \\ 
h_1 v_2 & h_2 \omega v_2 & h_3 \omega^2 v_2 \\ 
h_1 v_3 & h_2 \omega^2 v_3 & h_3 \omega v_3 \end{array} \right)
= \left( \begin{array}{ccc} 1 & 1 & 1 \\ 1 & \omega & \omega^2 \\ 
1 & \omega^2 & \omega \end{array} \right) \left( \begin{array}{ccc} 
h_1 v & 0 & 0 \\ 0 & h_2 v & 0 \\ 0 & 0 & h_3 v \end{array} \right),
\end{equation}
if $v_1=v_2=v_3=v$. This is the starting point of most subsequent A$_4$ 
models.  More recently, I discovered \cite{m06} that Eq.~(49) may also be 
obtained with $(\nu_i,l_i) \sim \underline{3}, ~l^c_i \sim \underline{3}$ and 
$(\phi^0_i,\phi^-_i) \sim \underline{1}, \underline{3}$, in which case
\begin{equation}
{\cal M}_l = \left( \begin{array}{ccc} h_0 v_0 & h_1 v_3 & h_2 v_2 \\ 
h_2 v_3 & h_0 v_0 & h_1 v_1 \\ h_1 v_2 & h_2 v_1 & h_0 v_0 \end{array} 
\right) = U_{l \nu}^{CW} \left( \begin{array}{ccc} m_e & 0 & 0 \\ 
0 & m_\mu & 0 \\ 0 & 0 & m_\tau \end{array} \right) (U_{l \nu}^{CW})^\dagger,
\end{equation}
if $v_1=v_2=v_3=v$.  Either way, $U_{l \nu}^{CW}$ has been derived.  To 
obtain $U_{l \nu}^{HPS}$, let ${\cal M}_\nu$ be Majorana and come from Higgs 
triplets: $(\xi^{++},\xi^+,\xi^0)$, then \cite{m04}
\begin{equation}
{\cal M}_\nu = \left( \begin{array}{ccc} a+b+c & f & e \\ f & a+\omega b + 
\omega^2 c & d \\ e & d & a + \omega^2 b + \omega c \end{array} \right),
\end{equation}
where $a$ comes from \underline{1}, $b$ from \underline{1}$'$, $c$ from 
\underline{1}$''$, and $(d,e,f)$ from \underline{3}.  To obtain Eq.~(50), 
we simply let $b=c$ and $e=f=0$.  Note that the tribimaximal mixing matrix 
does not depend on the neutrino mass eigenvalues $a-b+d$, $a+2b$, 
$-a+b+d$, nor the charged-lepton masses.  This implies the existence of 
residual symmetries \cite{l07,bhl08}.

Since \underline{1}$'$ and \underline{1}$''$ are unrelated in A$_4$, the 
condition $b=c$ is rather {\it ad hoc} .  A very clever solution was 
proposed by Altarelli and Feruglio \cite{af05}: they eliminated both 
\underline{1}$'$ and \underline{1}$''$ so that $b=c=0$. In that case, 
$m_1=a+d$, $m_2=a$, $m_3=-a+d$.  This is the simplest model of 
tribimaximal mixing, with the prediction of normal ordering of neutrino 
masses and the sum rule \cite{m05}
\begin{equation}
|m_{\nu_e}|^2 \simeq |m_{ee}|^2 + \Delta m^2_{atm}/9.
\end{equation}
Babu and He \cite{bh05} proposed instead to use 3 heavy neutral singlet 
fermions with ${\cal M}_D$ proportional to the identity and ${\cal M}_N$ 
of the form of Eq.~(50) with $b=0$.  In that case, the resulting 
${\cal M}_\nu$ has $b=c$ and $d^2=3b(b-a)$.  This scheme allows both 
normal and inverted ordering of neutrino masses.

The technical challenge in all such models is to break A$_4$ spontaneously 
along 2 incompatible directions: (1,1,1) with residual symmetry Z$_3$ in the 
charged-lepton sector and (1,0,0) with residual symmetry Z$_2$ in the neutrino 
sector. There is also a caveat.  If $\nu_2 = (\nu_e + \nu_\mu + \nu_\tau)/
\sqrt{3}$ remains an eigenstate, i.e. $e=f=0$, but $b \neq c$ is allowed, 
then the bound $|U_{e3}| < 0.16$ implies \cite{m04} $0.5 < \tan^2 \theta_{12} 
< 0.52$, away from the preferred experimental value of $0.45 \pm 0.05$.

\section{Beyond A$_4$ [S$_4$, $\Delta$(27), $\Sigma$(81), Q(24)]}

The group of permutation of 4 objects is S$_4$.  It contains both S$_3$ 
and A$_4$.  However, since the \underline{1}$'$ and \underline{1}$''$ 
of A$_4$ are now combined into the \underline{2} of S$_4$, tribimaximal 
mixing is achieved only with Eq.~(54). Furthermore, $h_1 \neq h_2$ in 
${\cal M}_l$ now requires both \underline{3} and \underline{3}$'$ 
Higgs representations.  No advantage appears to have been gained.

The group $\Delta(27)$ has the interesting decomposition $\underline{3} \times 
\underline{3} = \underline{\bar{3}} + \underline{\bar{3}} + 
\underline{\bar{3}}$, which allows
\begin{equation}
{\cal M}_\nu = \left( \begin{array}{ccc} x & fz & fy \\ fz & y & fx \\ 
fy & fx & z \end{array} \right).
\end{equation}
Using $\tan^2 \theta = 0.45$ and $\Delta m^2_{atm} = 2.7 \times 10^{-3}$ 
eV$^2$, this implies \cite{m08} $m_{ee} = 0.14$ eV.

The subgroups $\Sigma(3n^3)$ of U(3) may also be of interest.  $\Sigma(81)$ 
has 17 irreducible representations and may be applicable \cite{m07-2} 
to the Koide lepton mass formula
\begin{equation}
m_e + m_\mu + m_\tau = (2/3)(\sqrt{m_e} + \sqrt{m_\mu} + \sqrt{m_\tau})^2,
\end{equation}
as well as neutrino tribimaximal mixing \cite{m07-3}.

Since A$_4$ is a subgroup of SO(3), it has a spinorial extension which is 
a subgroup of SU(2).  This is the binary tetrahedral group, which has 24 
elements with 7 irreducible representations: \underline{1}, \underline{1}$'$, 
\underline{1}$''$, \underline{2}, \underline{2}$'$, \underline{2}$''$, 
\underline{3}.  It is also isomorphic to the quaternion group Q(24) whose 
24 elements form the vertices of the self-dual hyperdiamond in 4 dimensions. 
There have been several recent studies \cite{fhlm07,cm07,fk07,a07} 
involving Q(24), which may be useful for extending the success of A$_4$ 
for leptons to the quark sector. Note the peculiar fact that $A_4$ is not 
a subgroup of $Q(24)$.

\begin{acknowledgments}
This work was supported in part by the U.~S.~Department of Energy under Grant 
No.~DE-FG03-94ER40837.  I thank Meltem Serin for her great hospitality at 
Akyaka.
\end{acknowledgments}

\end{document}